\documentclass[12pt]{article}
\pdfoutput=1
\usepackage{amsmath,amssymb,amsfonts,amsthm,bm,bbm,cancel,wasysym}
\usepackage{epsfig,graphics,graphicx,epstopdf,caption,subcaption}
\usepackage[numbers,sort&compress]{natbib}
\graphicspath{{Charts/}}
\usepackage{array,booktabs,colortbl,colordvi,multirow}
\usepackage{colordvi,color,xcolor}
\usepackage{hyperref}
\usepackage{rotating}
\usepackage{comment}
\usepackage{feynmf}

\usepackage[symbol]{footmisc}
\renewcommand{\thefootnote}{\fnsymbol{footnote}}

\begin{document}

\begin{center}

{\Large {\bf Gravitational Wave Probe of Gravitational Dark Matter from Preheating}}\\

\vspace*{0.75cm}

{Ruopeng Zhang and Sibo Zheng\footnote{Corresponding author: sibozheng.zju@gmail.com.}}

\vspace{0.5cm}
{Department of Physics, Chongqing University, Chongqing 401331, China}
\end{center}
\vspace{.5cm}

\begin{abstract}
\noindent
We forecast high-frequency gravitational wave (GW)  from preheating hosting gravitational dark matter (GDM) as the indirect probe of such GDM.
We use proper lattice simulations to handle resonance, 
and to solve GW equation of motion with the resonance induced scalar field excitations as source term. 
Our numerical results show that Higgs scalar excitations in Higgs preheating model give rise to magnitudes of GW energy density spectra of order $10^{-10}$ at frequencies $10-10^{3}$ MHz depending on the GDM mass of $(6-9)\times 10^{13}$ GeV,
whereas inflaton fluctuation excitations in inflaton self-resonant preheating model yield magnitudes of GW energy density spectrum up to $10^{-9}~(10^{-11})$ at frequencies near $30~(2)$ MHz for the index $n=4~(6)$ with respect to the GDM mass of $1.04~(2.66)\times 10^{14}$ GeV.
\end{abstract}

\renewcommand{\thefootnote}{\arabic{footnote}}
\setcounter{footnote}{0}
\thispagestyle{empty}
\vfill
\newpage
\setcounter{page}{1}

\tableofcontents

\section{Introduction}
After the end of inflation the early Universe is believed to evolve with time in a manner such that the inflaton energy density is transferred to the radiation energy density.
No matter how the transit is achieved, this epoch provides a circumstance where various particles including dark matter (DM) can be produced.
Motivated by null results of experiments aiming to detect non-gravitational DM,
in this work we focus on gravitational dark matter (GDM) that only interacts with other matters via the gravitational portal, 
see \cite{Kolb:2023ydq} for a recent review. 

GDM can be produced in the early Universe either in the circumstance of reheating \cite{Markkanen:2015xuw,Fairbairn:2018bsw,Hashiba:2018tbu,Ema:2019yrd,Ahmed:2020fhc,Babichev:2020yeo, Mambrini:2021zpp,Barman:2021ugy,Clery:2021bwz,Barman:2022qgt,Lebedev:2022vwf,Tang:2017hvq,Garny:2017kha,Chianese:2020yjo} or recently studied preheating  \cite{Karam:2020rpa,Klaric:2022qly,Zhang:2023xcd,Zhang:2023hjk}.
In the case of reheating, a perturbative ``decay" of the parent inflaton condensate converts its energy density into the daughter Standard Model (SM) particles,
whereas in the case of preheating this energy transfer is achieved via a resonant production of daughter fields caused by oscillations of the parent inflaton condensate.
While these studies have demonstrated that the observed DM relic abundance can be satisfied, 
it is rather challenging to test GDM due to the following features.
First, GDM is absolutely stable, implying that indirect limits derived from DM decay are not viable.
Second, either GDM annihilation, self scattering, or scattering off SM particles are always Planck-scale suppressed,
which are out of reaches of these experiments. 
Compared to these late-time measurements, other early-time measurements seem more promising to test the GDM.
This is the subject of this work.

\newpage
In this study we forecast high-frequency gravitational wave (GW)  from preheating hosting GDM as the indirect probe of such GDM.
In the context of reheating, GWs are emitted by perturbative decay induced daughter fields among others \cite{Garcia-Bellido:2007fiu, Bernal:2019lpc,Ghoshal:2022ruy,Barman:2023ktz,Choi:2024ilx}.
Likewise, GWs arising from preheating are emitted by resonance induced daughter fields \cite{Khlebnikov:1997di, Easther:2006vd, Dufaux:2007pt, Dufaux:2008dn,Dufaux:2010cf,Zhou:2013tsa, Bethke:2013vca, Antusch:2016con,Figueroa:2017vfa, Antusch:2017vga, Amin:2018xfe, Adshead:2018doq, Lozanov:2019ylm, Adshead:2019lbr, Krajewski:2022ezo, Cosme:2022htl, Garcia:2023eol}.
Once created, these GWs immediately decouple on the contrary to the Cosmic Microwave Background (CMB).
Notably, GWs from preheating are determined by relevant resonance parameter controlling the source term of GW equation. 
The resonant parameter can be fixed if the preheating is required to address the observed GDM relic density as shown by our previous studies in \cite{Zhang:2023xcd,Zhang:2023hjk}, where an explicit one-to-one correspondence between the resonant parameter and GDM mass has been uncovered in the individual preheating model.
In this situation these GWs can be promoted to serve as the indirect probe of GDM.
On the contrary, if the correspondence is obscure or absent, they are at best the indirect probe of preheating as previously considered in the literature.

The rest of the paper is organized as follows. 
Sec.\ref{HR} addresses the GW production in the Higgs preheating.
We firstly discuss the resonant excitations of Higgs field, then use the obtained results to derive the present-day GW energy density spectra, and compare them to other parameter resonance induced GWs. 
Sec.\ref{IF} addresses the GW production in the minimal preheating.
We firstly analyze the self-resonant excitations of inflaton fluctuation, 
then derive the present-day GW energy density spectra as in Sec.\ref{HR},
and explain how these GWs differ from those in the literature.
To carry out the GW spectra in each preheating model, we use proper lattice simulation \cite{Figueroa:2020rrl,Figueroa:2021yhd,Cosmolattice} to handle the resonant process.
We will  clearly  show the correlation between the growth of GWs and the excitations of relevant daughter scalar field.
Appendix.\ref{evalGW} shows how to evaluate the GW spectra from the end time of preheating to the present time.
Finally, we conclude in Sec.\ref{con}, where we briefly comment on detection prospects of the derived GW spectra.

Note: throughout the text $M_{P}=2.4\times 10^{18}$ GeV is the reduced Planck mass scale, the subscript ``end" and ``$_{*}$"  denote the end of inflation and of lattice simulation respectively, and the subscript ``0" is the present-day value. 

\section{The Higgs preheating}
\label{HR}
The Higgs preheating is built upon the following Lagrangian 
\begin{eqnarray}{\label{Lag1}}
 \mathcal{L}=\frac{1}{2} \partial_{\mu}\phi\partial^{\mu}\phi+\left(D_{\mu}H\right)^{\dag}D^{\mu}H-V(\phi)-\kappa\mid H\mid^{2}\phi^{2},
\end{eqnarray}
where $V(\phi)$ denotes the inflation potential and the last term represents the interaction between the inflaton $\phi$ and the SM Higgs doublet with $\kappa$ being the dimensionless coupling constant.
In \cite{Zhang:2023xcd} we have specifically considered the $\alpha$-attractor T-model of inflation \cite{Kallosh:2013hoa,Kallosh:2013yoa} with
\begin{eqnarray}{\label{V1}}
V(\phi)=\lambda_{\phi} M_{P}^{4}\left[\sqrt{6}\tanh\left(\frac{\phi}{\sqrt{6}M_{P}}\right)\right]^{2} \rightarrow \frac{1}{2}m^{2}_{\phi}\phi^{2},
\end{eqnarray}
which is approximated by a quadratic mass term with $m_{\phi}=\sqrt{2\lambda_{\phi}}M_{P}$ in the field range of $\mid\phi\mid<< M_{P}$.

Ignoring the Higgs self-interaction\footnote{The magnitude of Higgs self-coupling has to be small enough \cite{Zhang:2023xcd} in the ultraviolet energy region for the parameter resonance to occur. Such small Higgs self-coupling can be achieved by modifying its renormalization group equation as discussed in the context of SM electroweak vacuum stability.}, we place constraints on the model parameters in eq.(\ref{Lag1}) as follows. 
First, one uses the Planck 2018 data \cite{Planck:2018jri} to fix the value of $\lambda_{\phi}$ in eq.(\ref{V1}) as $\lambda_{\phi}\approx 2.05\times 10^{-11}$.
Second, the coupling $\kappa$ is upper bounded by the flatness of inflation potential  as $\kappa<\sqrt{\lambda_{\phi}}$ 
and lower bounded by the preheating condition as $\kappa>\kappa_{\rm{min}}$, 
with $\kappa_{\rm{min}}\approx 2.5\times 10^{-7}$ the minimal value of $\kappa$ required by the parameter resonance.
So we are left with the single model parameter $\kappa$ constrained to be in the narrow range of
\begin{eqnarray}{\label{kappa}}
\kappa\sim (3-13)\times 10^{-7}.
\end{eqnarray}
Within this resonant parameter range, the GDM mass $m_{\chi}\sim (4-6)m_{\phi}\sim (6-9)\times 10^{13}$ GeV due to the one-to-one correspondence \cite{Zhang:2023xcd} between $\kappa$ and $m_{\chi}$ imposed by the observed DM relic density.

\subsection{Lattice treatment on parameter resonance}
We begin with the equations of motion for the inflaton condensate and the SM Higgs:
\begin{eqnarray}\label{eom1}
\left(\frac{d^{2}}{dt^{2}}+3H\frac{d}{dt}\right)\phi+\frac{\partial V(\phi)}{\partial \phi} &=& 0, \nonumber\\
\left(\frac{d^{2}}{dt^{2}}-\frac{\nabla^{2}}{a^{2}}+3H\frac{d}{dt}+m^{2}_{h,\lambda}\right)h_{\lambda}&=&0, 
\end{eqnarray}
where $h_{\lambda}$ are the four real scalars in the Higgs doublet with $\lambda=1$-4, $H$ is the Hubble rate, $a$ is the scale factor, ``$\nabla$" is derivative over space coordinates, 
and $m^{2}_{h,\lambda}\approx \kappa\phi^{2}$ is the effective mass squared varying with time due to the oscillation of $\phi$ after the end of inflation.

To understand the Higgs field excitation due to the parameter resonance, 
we make use of lattice simulations. 
In the lattice simulation,  it is more convenient to introduce the re-scaled field variables instead of those in eq.(\ref{eom1})
\begin{eqnarray}\label{rescale}
\tilde{h}_{\lambda}=\frac{h_{\lambda}}{\phi_{\rm{end}}}~~\rm{and}~~\tilde{\phi}=\frac{\phi}{\phi_{\rm{end}}},
\end{eqnarray}
where $\phi_{\rm{end}}$ is the initial value of inflaton field after inflation. 
With the initial conditions \cite{Zhang:2023xcd},
\begin{eqnarray}\label{initial1}
\rho_{\phi,\rm{end}}=7.1\times 10^{62}\rm{GeV}^{4},~~
\rho_{H,\rm{end}}=0,~~
\phi_{\rm{end}}= 0.84~M_{P},
\end{eqnarray}
we solve eq.(\ref{eom1}) in terms of publicly available lattice code CosmoLattice \cite{Figueroa:2020rrl,Figueroa:2021yhd},
by adopting the lattice size $N=128^{3}$ and the minimal infrared cut-off $k_{IR}=1$ in the following discussions of this section.
In this code the Hubble parameter is solved self-consistently through Friedmann equation.

\begin{figure}
\centering
\includegraphics[width=8cm,height=8cm]{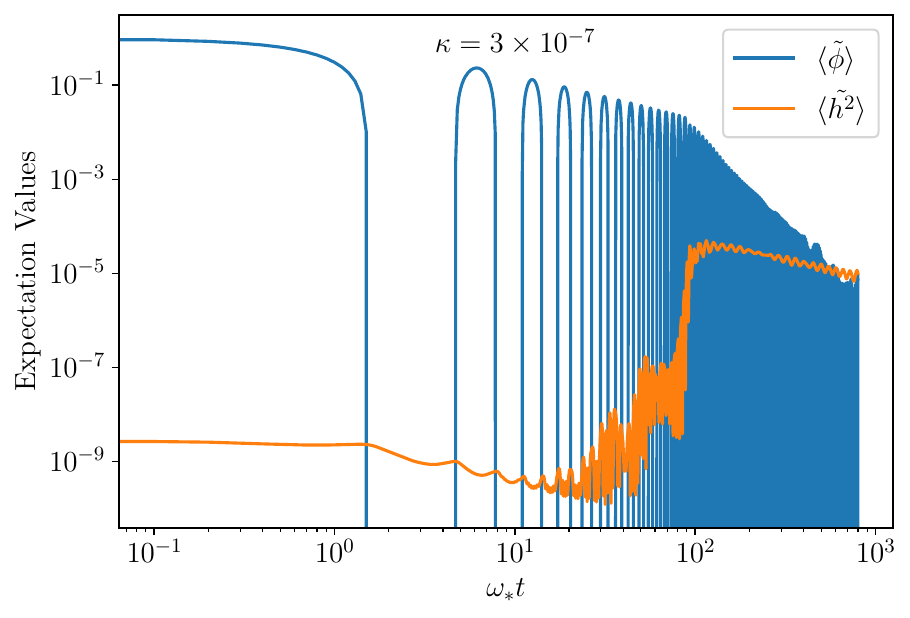}
\includegraphics[width=8cm,height=8cm]{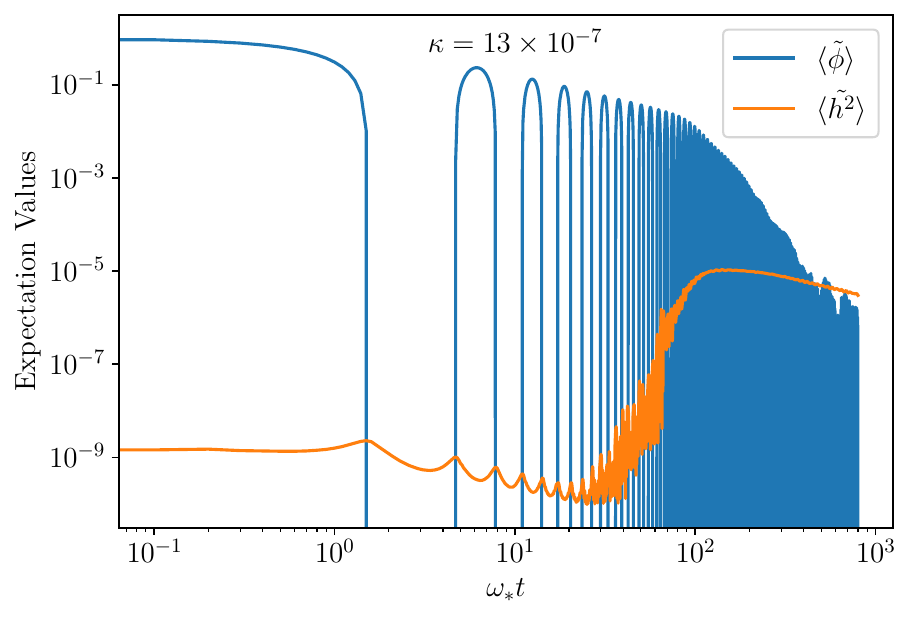}\\
\includegraphics[width=8cm,height=8cm]{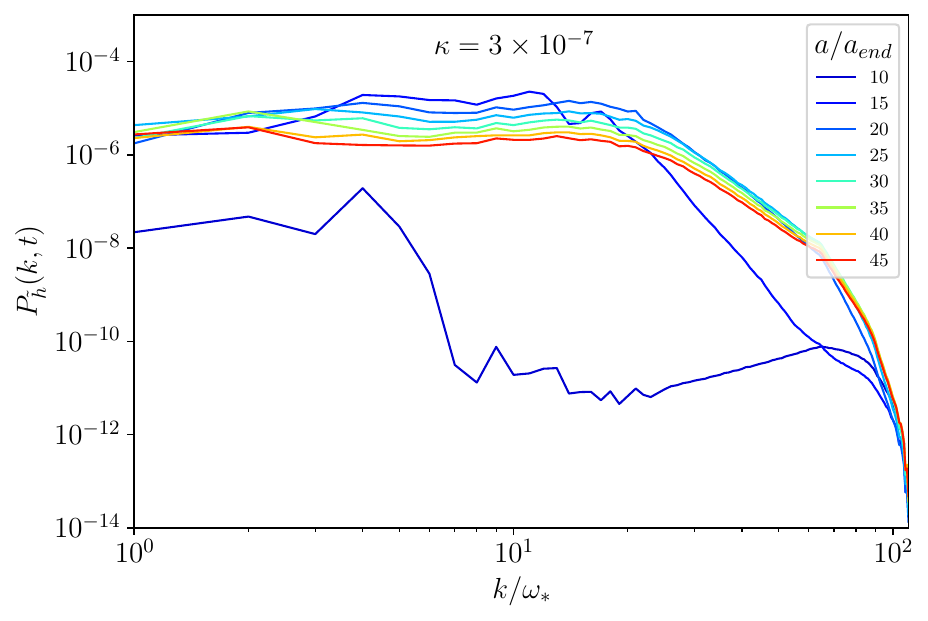}
\includegraphics[width=8cm,height=8cm]{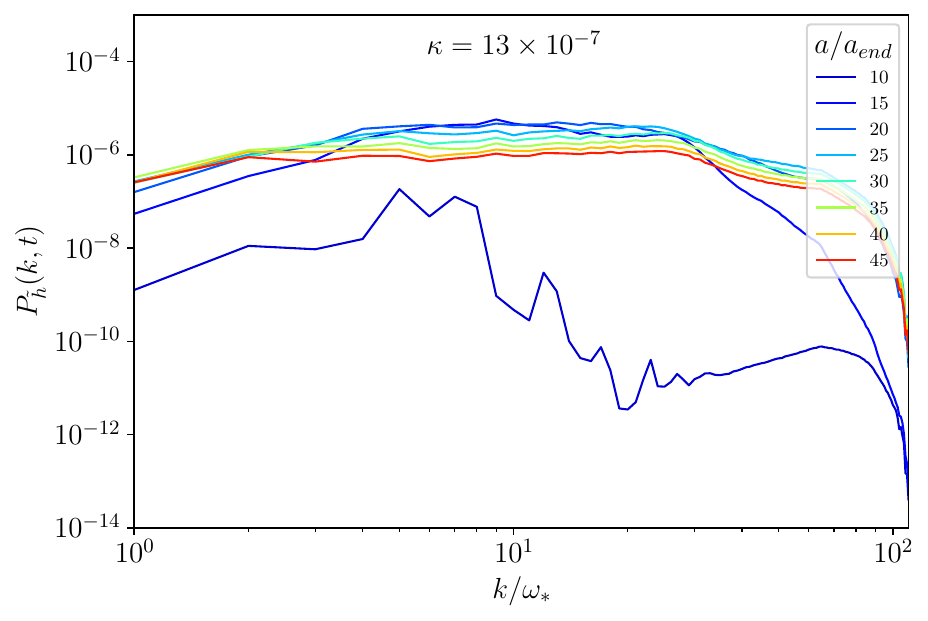}
\centering
\caption{$\mathbf{Top}$: the evolution of inflaton's  volume average $\left<\tilde{\phi}\right>$ (in blue) and Higgs variance $\left<\tilde{h}^{2}_{\lambda}\right>$ (in orange) as function of time in units of $\omega_{*}$ for $\kappa=3\times10^{-7}$ (left) and $\kappa=13\times10^{-7}$ respectively. $\mathbf{Bottom}$: the evolution of Higgs power spectrum $P_{\tilde{h}}$ as function of $k$ with respect to time for $\kappa=3\times10^{-7}$ (left) and $\kappa=13\times10^{-7}$ respectively.}
\label{hvar}
\end{figure}

The top panel of fig.\ref{hvar} presents the evolution of the inflaton's volume average $\left<\tilde{\phi}\right>$ (in blue) and the Higgs variance $\left<\tilde{h}^{2}_{\lambda}\right>$ (in orange) as function of time (in units of $\omega_{*}=m_{\phi}$) for $\kappa=3\times 10^{-7}$ (left) and $\kappa=13\times 10^{-7}$ (right) respectively.
In this panel it is clear to see two critical points. 
(i) After a few oscillations, whenever the inflaton condensate crosses the zero points, i.e, $\tilde{\phi}(t)\rightarrow 0$, the value of $\left<\tilde{h}^{2}_{\lambda}\right>$ rapidly grows, as the effective mass $m^{2}_{h,\lambda}$ in eq.(\ref{eom1}) vanishes at these points.
(ii) When the backreaction due to the Higgs field becomes strong, which corresponds to the time $\omega_{*}t=100~(150)$ for $\kappa=3~(13)\times 10^{7}$,
 the growth of the Higgs variance stops.

The bottom panel of fig.\ref{hvar} shows the evolution of Higgs power spectrum defined by 
\begin{eqnarray}\label{powers}
\left<\tilde{h}_{\lambda}(\mathbf{k})\tilde{h}^{*}_{\lambda}(\mathbf{k'})\right>=(2\pi)^{3}P_{\tilde{h}_{\lambda}}(k)\delta(\mathbf{k}-\mathbf{k'}),
\end{eqnarray}
with respect to time for $\kappa=3\times 10^{-7}$ (left) and $\kappa=13\times 10^{-7}$ (right) respectively, 
with the comoving wavenumber $k$ in units of $\omega_{*}$.
Comparing the top and bottom panel, with the end of simulation time $a(t_{*})/a_{\rm{end}}=45$ in the later one corresponding to $\omega_{*}t_{*}\approx 720$ in the former one, 
one clearly sees that the growth of $P_{\tilde{h}_{\lambda}}(k)$ stops whenever the value of $\left<\tilde{h}^{2}_{\lambda}\right>$ gets stabilized.
In addition, comparing the two red curves of $P_{\tilde{h}}$ in the bottom panel, 
one finds that as $\kappa$ increases, the value of $P_{\tilde{h}}$ increases in the large $k$ regime, 
but it becomes smaller in the small $k$ regime. 
This feature will help us understand the dependence of GW spectra on $\kappa$  as shown in fig.\ref{hGWf}.

\subsection{Gravitational wave production}
Having addressed the Higgs field excitations, we now derive the GWs produced during the resonant process. 
The GWs are determined by the equation of motion for the tensor perturbation $h_{ij}$ as 
\begin{eqnarray}\label{GWe1}
\ddot{h}_{ij}+3H\dot{h}_{ij}-\frac{\nabla^{2}}{a^{2}}h_{ij}=\frac{2}{M^{2}_{P}a^{2}}\Pi^{TT}_{ij},
\end{eqnarray}
where $\Pi^{TT}_{ij}=\Lambda_{ijk\ell}\Pi_{k\ell}$ is the transverse-traceless part of the effective anisotropic stress tensor,  
with 
\begin{eqnarray}\label{GWs1}
\Pi_{k\ell}=\partial_{k}\phi\partial_{\ell}\phi+\sum_{\lambda}\partial_{k}h_{\lambda}\partial_{\ell}h_{\lambda},
\end{eqnarray}
and $\Lambda_{ijk\ell}$ the projector.
In terms of eq.(\ref{GWe1}) one obtains the produced GW energy density spectrum 
\begin{eqnarray}\label{GW1}
\Omega_{\rm{GW}}(k,t)=\frac{1}{\rho_{c}}\frac{d\rho_{\rm{GW}}}{d\log k}(k, t)=\frac{k^{3}}{(4\pi)^{3}GV}\int \frac{d\Omega_{k}}{4\pi}\dot{h}_{ij}(\mathbf{k}, t)\dot{h}^{*}_{ij}(\mathbf{k}, t),
\end{eqnarray}
where $\rho_c$ is the total energy density during the preheating, $\Omega_{k}$ is the solid angle of Fourier space $\mathbf{k}$, and $V$ is the volume. 

We use the lattice simulations developed in CosmoLattice \cite{Cosmolattice} to solve a discrete version of eq.(\ref{GWe1}) in the Fourier space.
Instead of projecting $\Pi$ to obtain $\Pi^{TT}$ and then solving $h_{ij}$ at every time step of evolution, 
in practice this lattice code directly uses $\Pi$ to derive a tensor perturbation $u_{ij}$ and then imposes the $\Lambda$ operator on $u_{ij}$ to obtain the physical $h_{ij}$. 
By doing so the consumption of simulation time is significantly reduced while the validity is guaranteed as both operations commute with each other. 
We refer the reader to \cite{Cosmolattice} for more details about this point. 

\begin{figure}
\centering
\includegraphics[width=8cm,height=8cm]{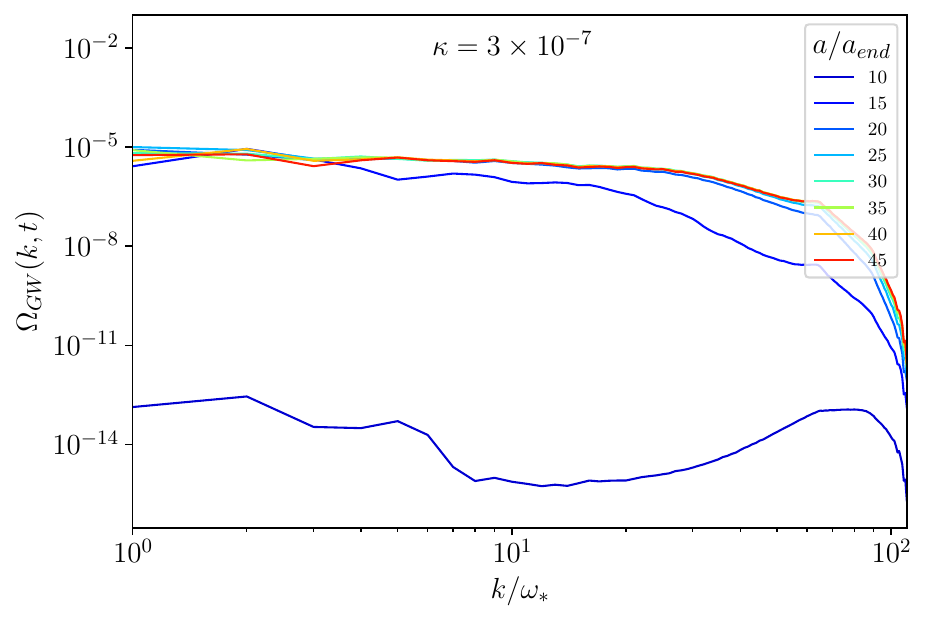}
\includegraphics[width=8cm,height=8cm]{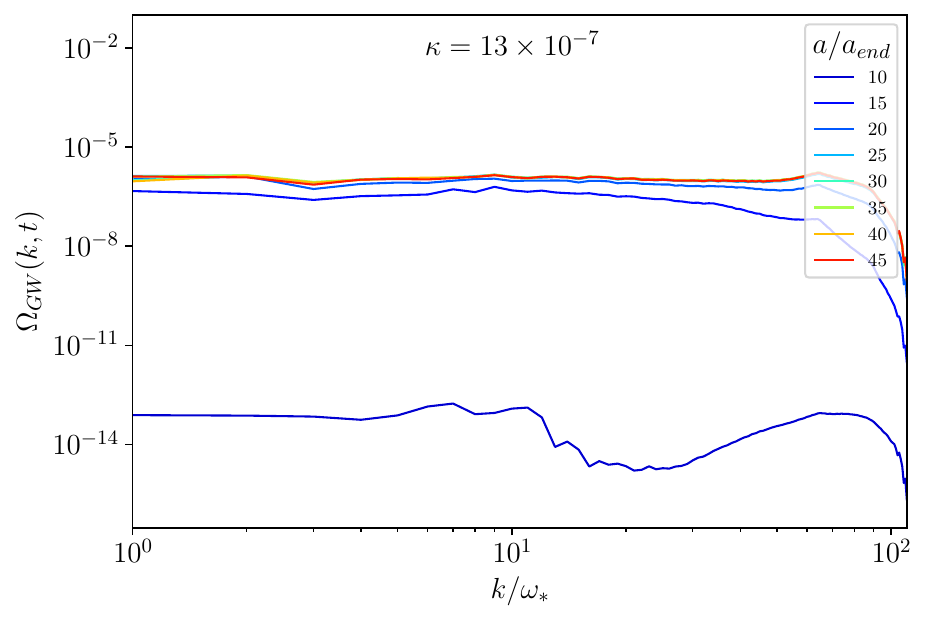}
\centering
\caption{The Higgs excitations induced GW spectra with respect to time during the Higgs preheating for $\kappa=3\times10^{-7}$ (left) and $\kappa=13\times10^{-7}$ (right) respectively, 
where the time indices of the color curves are the same as in the bottom panel of fig.\ref{hvar}. }
\label{hGW}
\end{figure}

Fig.\ref{hGW} shows the GW energy density spectra in eq.(\ref{GW1}) with respect to time for the two representative values of $\kappa=3\times 10^{-7}$ (left) and $\kappa=13\times 10^{-7}$ (right) respectively. 
We understand the evolution of the GW spectra by comparing fig.\ref{hGW} to the bottom panel of fig.\ref{hvar}, where the time indices are the same.
One finds that the time evolution of $\Omega_{\rm{GW}}(k)$ is correlated to that of $P_{\tilde{h}}(k)$,
in the sense that the growth of the GW spectra almost immediately stops once the growth of $P_{\tilde{h}}(k)$ stops as shown by the red curves in these two figures.
Such correlation between the resonance induced daughter field excitations and the daughter field induced GWs  has also been verified by the study of another preheating in \cite{Cosme:2022htl}.

In terms of the GW spectra at the end time of stimulation, with $a(t_{*})/a_{\rm{end}}=45$ as shown in fig.\ref{hGW},
we show the present-day GW spectra $\Omega_{\rm{GW},0}(f)$ with respect to various values of $\kappa$ in the left panel of fig.\ref{hGWf},
by evaluating $\Omega_{\rm{GW}}(k, t_{*})$ from $t_{*}$ to the present time $t_{0}$ as shown in the appendix.\ref{evalGW}. 
In this figure the dependence of $\Omega_{\rm{GW},0}(f)$ on the model parameter $\kappa$ has been replaced by that of $\Omega_{\rm{GW},0}(f)$ on the GDM mass $m_{\chi}$, 
using the one-to-one correspondence between $\kappa$ and $m_{\chi}$ \cite{Zhang:2023xcd}.
As $\kappa$ (alternatively $m_{\chi}$) increases, the magnitude of $\Omega_{\rm{GW},0}(f)$ is larger in the large $f$ regime,
but it is smaller in the small $f$ regime, as previously seen in the dependence of $P_{\tilde{h}_{\lambda}}(k)$ on $\kappa$ in fig.\ref{hvar}.
Compared to the parameter resonance induced GWs studied by \cite{Bethke:2013vca, Figueroa:2017vfa},
the Higgs resonance induced GWs spectra in fig.\ref{hGWf} have relatively smaller peak frequencies and larger peak energy densities.

The emitted GWs can be understood as radiation, whose contribution to the effective neutrino number is determined by
 \begin{eqnarray}\label{GWtot1}
\frac{7}{8}\left(\frac{4}{11}\right)^{4/3}\Delta N_{\rm{eff}}=\frac{\Omega^{\rm{tot}}_{\rm{GW},0}}{\Omega_{\rm{rad},0}},
\end{eqnarray}
where the total GW energy density
 \begin{eqnarray}\label{GWtot11}
\Omega^{\rm{tot}}_{\rm{GW},0}=\int d\log k~\Omega_{\rm{GW}}(k,t_{0}).
\end{eqnarray}
Substituting the results of the left panel of fig.\ref{hGWf} into eq.(\ref{GWtot1}), one obtains the effective neutrino number as shown in the right panel of fig.\ref{hGWf}.
Therein the values of $\Delta N_{\rm{eff}}$ range from $\sim 1\times 10^{-5}$ to $4\times10^{-5}$, depending on the GDM mass $m_{\chi}$.

\begin{figure}
\centering
\includegraphics[width=8cm,height=8cm]{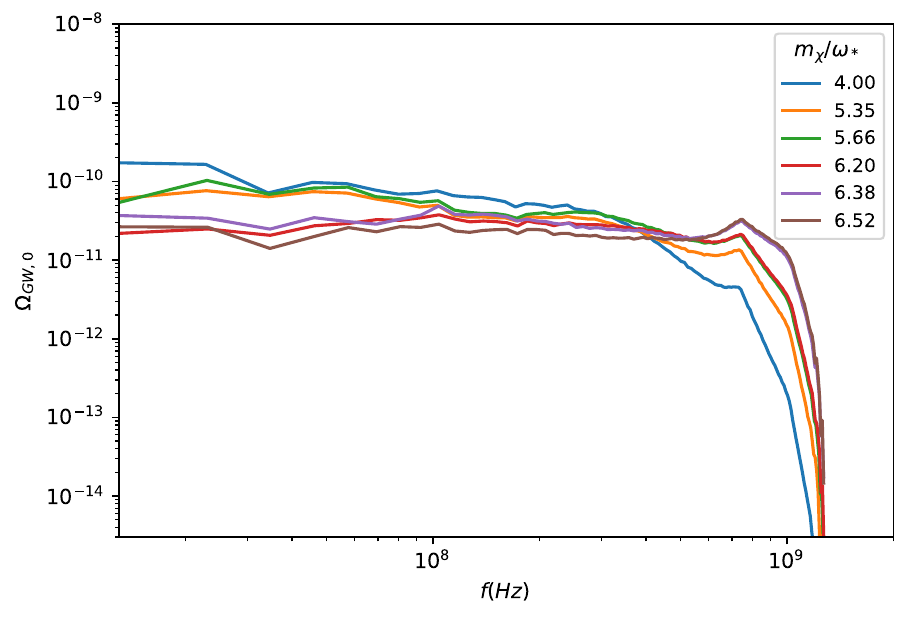}
\includegraphics[width=8cm,height=8cm]{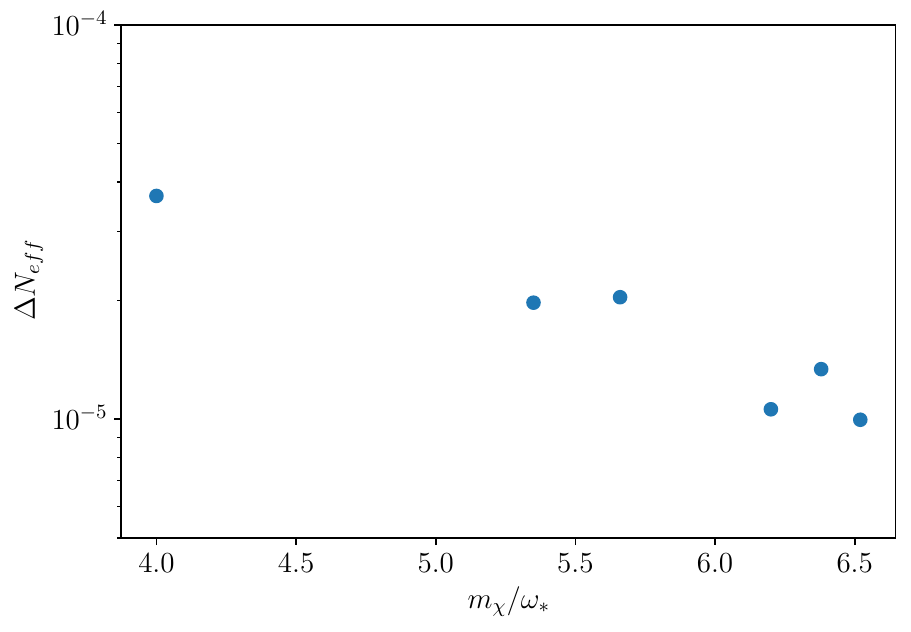}
\centering
\caption{$\mathbf{Left}$: the present-day GW spectra with respect to various values of GDM mass $m_{\chi}$ using the one-to-one correspondence between $\kappa$ and $m_{\chi}$ \cite{Zhang:2023xcd}. $\mathbf{Right}$: the individual GW contribution to the effective neutrino number.}
\label{hGWf}
\end{figure}

\section{The minimal preheating}
\label{IF}
Unlike the Higgs preheating where the coupling of inflaton to the SM Higgs has to be introduced, 
the Lagrangian of the minimal preheating simply reads as 
\begin{eqnarray}{\label{Lag2}}
 \mathcal{L}=\frac{1}{2} \partial_{\mu}\phi\partial^{\mu}\phi-V(\phi)
\end{eqnarray}
where the inflation potential $V(\phi)$ is given by \cite{Kallosh:2013hoa,Kallosh:2013yoa}
\begin{eqnarray}{\label{V2}}
V(\phi)=\lambda_{\phi} M_{P}^{4}\left[\sqrt{6}\tanh\left(\frac{\phi}{\sqrt{6}M_{P}}\right)\right]^{n} \rightarrow \lambda_{\phi} \frac{\phi^{n}}{M^{n-4}_{P}},
\end{eqnarray}
with the index $n$ being an even number in order to make sure the positivity of the inflation potential. 
As explained in the beginning of Sec.\ref{HR}, the model parameter $\lambda_\phi$ has been fixed by the Planck data,
implying that $n$ is the only free parameter in eq.(\ref{Lag2}).
For the specific situation with $n=4~(6)$, the GDM mass $m_{\chi}\approx 1.04~(2.66)\times 10^{14}$ GeV 
due to the one-to-one correspondence  \cite{Zhang:2023hjk} between $m_{\chi}$ and $n$ imposed by the observed DM relic density.

\subsection{Lattice treatment on self-resonance}
Similar to eq.(\ref{eom1}) we start with the equations of motion for the inflaton condensate and its fluctuation $\delta\phi$:
\begin{eqnarray}{\label{eom2}}
\ddot{\phi}+3H\dot{\phi}+\frac{\partial V}{\partial \phi} &=& 0, \nonumber\\
\ddot{\delta\phi}+3H\dot{\delta\phi}-\frac{\nabla^{2}}{a^{2}}\delta\phi+\frac{\partial^{2}V}{\partial \phi^{2}}\delta\phi&=&0.
\end{eqnarray}
Previously, the results of \cite{Amin:2010dc,Amin:2011hj,Lozanov:2016hid,Lozanov:2017hjm} have shown that 
a self-resonant excitation of the inflaton fluctuations takes place for $n=4$ and 6,
which has been verified by our recent work \cite{Zhang:2023hjk}.
One can understand the self-resonance as a mixture of both parameter and tachyonic resonance by expanding the inflation field as $\phi(t)+\delta\phi(\mathbf{x},t)$ in eq.(\ref{V2}).
In this sense, the analogy of the Higgs variance $\left<\tilde{\delta\phi}^{2}\right>$, with $\tilde{\delta\phi}=\delta\phi/\phi_{\rm{end}}$ defined as in eq.(\ref{rescale}), is no longer suitable to illustrate the self-resonance.
Rather, the analogy of Higgs power spectrum $P_{\tilde{\delta\phi}}(k)$ still works, 
because the growth of $P_{\tilde{\delta\phi}}(k)$ is tied to that of GWs emitted by the inflaton fluctuation excitations.

\begin{figure}
\centering
\includegraphics[width=8cm,height=8cm]{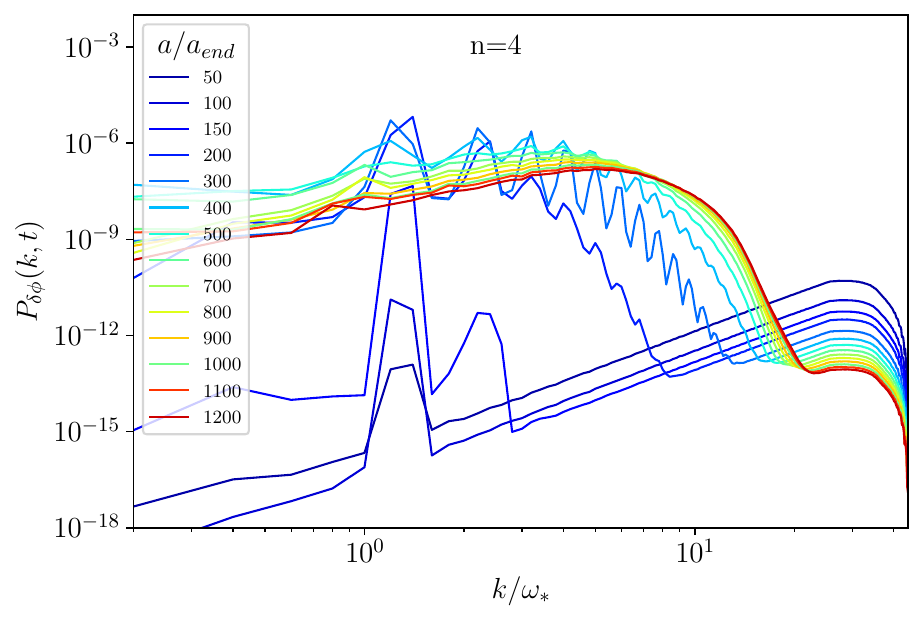}
\includegraphics[width=8cm,height=8cm]{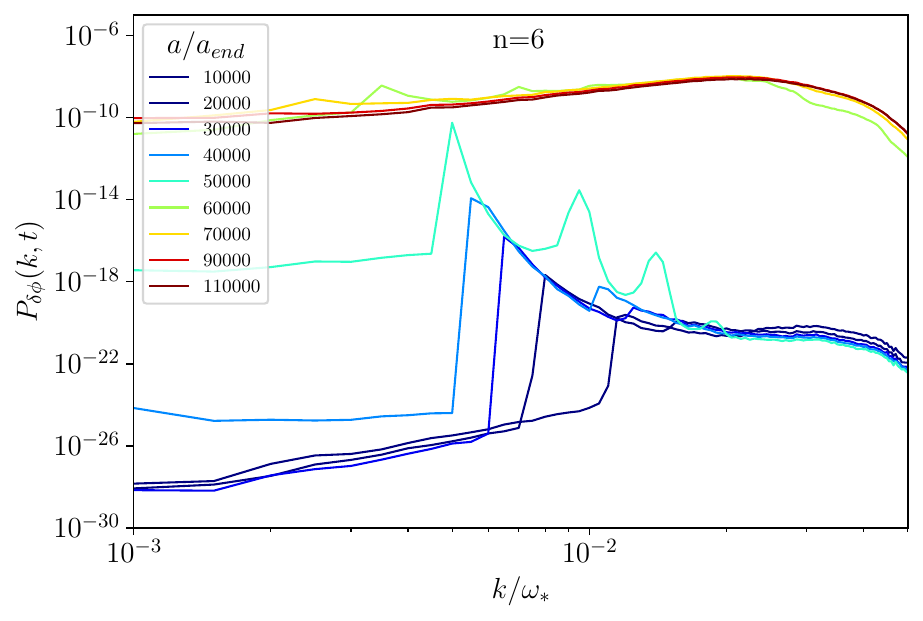}
\centering
\caption{The evolution of the power spectrum $P_{\tilde{\delta\phi}}(k)$ with respect to time for $n=4$ (left) and $n=6$ (right) respectively, where 
$k$ is in units of $\omega_{*}$.}
\label{mvar}
\end{figure}

Using the initial conditions displayed in Table 1 of \cite{Zhang:2023hjk} and adopting the lattice size $N=256^{3}~(128^{3})$ and the minimal infrared cut-off $k_{IR}=0.2~(5\times 10^{-4})$ for $n=4~(6)$\footnote{One has to choose different values of $k_{IR}$ in order to capture the different resonant bands with respect to different values of $n$.}
to solve eq.(\ref{eom2}),
we show the evolution of $P_{\tilde{\delta\phi}}(k)$ with respect to time in fig.\ref{mvar}, with $k$ in units of $\omega_{*}$.
The explicit values of $\omega_{*}$ can be found in \cite{Zhang:2023hjk}.
Similar to the parameter resonance induced Higgs power spectrum $P_{\tilde{h}}(k)$ in Sec.\ref{HR}, 
the self-resonance induced power spectrum $P_{\tilde{\delta\phi}}(k)$ in this figure rapidly grows until the backreaction becomes strong. 
Afterward, it becomes stabilized as shown by the red curves with respect to $a(t_{*})/a_{\rm{end}}\approx1.2\times 10^{3}~(1.1\times10^{5})$ for $n=4~(6)$ respectively. 
Compared to $n=4$ in the left plot, the resonant peaks for $n=6$ in the right plot are much more suppressed. 
The reason is that the oscillating frequency $\omega_{k}=(k^{2}+a^{2}\partial^{2}V/\partial\phi^{2})^{1/2}\sim a^{0}$ and $\sim a^{-1/2}$ for $n=4$ and 6 respectively, implying a relatively postponed growth of the inflaton fluctuations in the later case.

\subsection{Gravitational wave production}
Similar to eq.(\ref{GWe1}) the equation of motion for the tensor perturbation $h_{ij}$ during the minimal preheating is given by 
\begin{eqnarray}\label{GWe2}
\ddot{h}_{ij}+3H\dot{h}_{ij}-\frac{\nabla^{2}}{a^{2}}h_{ij}=\frac{2}{M^{2}_{P}a^{2}}\Pi^{TT}_{ij},
\end{eqnarray}
but with the effective anisotropic stress tensor 
\begin{eqnarray}\label{GWs2}
\Pi_{ij}=\partial_{i}\delta\phi\partial_{j}\delta\phi,
\end{eqnarray}
instead of eq.(\ref{GWs1}).

\begin{figure}
\centering
\includegraphics[width=8cm,height=8cm]{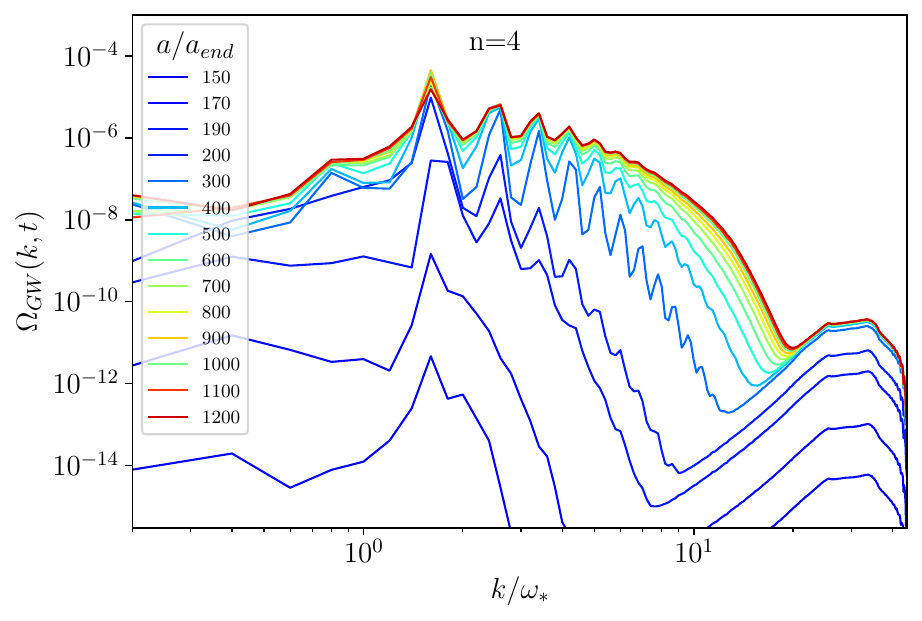}
\includegraphics[width=8cm,height=8cm]{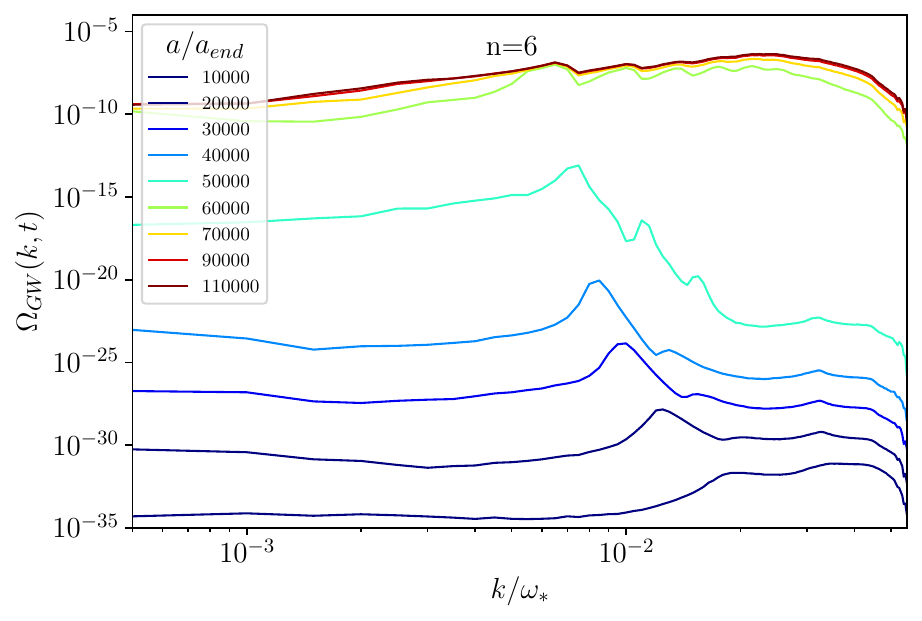}
\centering
\caption{The inflaton fluctuation excitations induced GW spectra during the minimal preheating with respect to time for $n=4$ (left) and $n=6$ (right) respectively, 
where the time indices of the color curves are the same as in fig.\ref{mvar}. }
\label{mGW}
\end{figure}

After solving the discrete version of eq.(\ref{GWe2}), we show in fig.\ref{mGW}  the evolution of GWs produced during the minimal preheating with respect to time for $n=4$ (left) and $n=6$ (right) respectively, 
where the time indices of the color curves are the same as in fig.\ref{mvar}.
Similar to the correlation observed from figs.\ref{hvar} and \ref{hGW} in the case of Higgs preheating, 
the growth of $\Omega_{\rm{GW}}(k)$ in fig.\ref{mGW} is correlated to that of $P_{\tilde{\delta\phi}}(k)$ in fig.\ref{mvar} as well in the case of minimal preheating either for $n=4$ or $n=6$.
Fig.\ref{mGW} shows that there are multiple local peak frequencies in the $k$ regime of interest.These local peaks turn out to be imprinted in the present-day GW spectra as below.

\begin{figure}
\centering
\includegraphics[width=12cm,height=8cm]{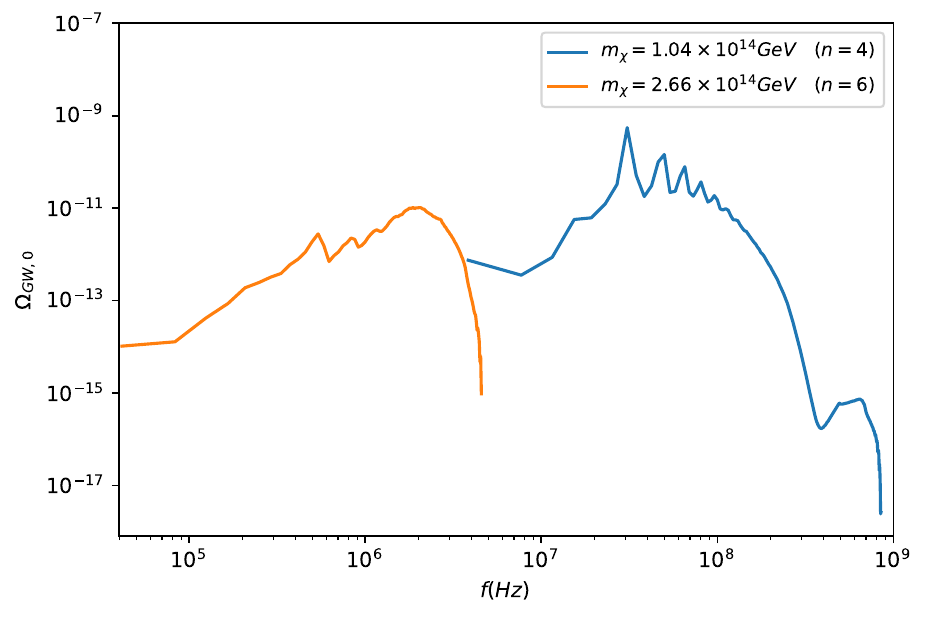}
\centering
\caption{The present-day GW spectra with respect to the GDM mass $m_{\chi}=1.04~(2.66)\times10^{14}$ GeV for $n=4~(6)$, using the one-to-one correspondence between $m_{\chi}$ and $n$ \cite{Zhang:2023hjk}.}
\label{mGWf}
\end{figure}

Using the results of fig.\ref{mGW} we show the present-day GW spectra in fig.\ref{mGWf}, 
by evaluating $\Omega_{\rm{GW}}(k)$ from the end time of preheating $t_{*}$ to the present time $t_{0}$ as shown in Appendix.\ref{evalGW}.
In this figure several local peak frequencies previously seen in fig.\ref{mGW} appear near $\sim 40~(1)$ MHz for $n=4~(6)$.
Moreover, the peak value of $\Omega_{\rm{GW},0}$ for $n=4$ is about two orders of magnitude larger than that for $n=6$, 
due to the relatively larger inflaton fluctuation excitations as seen in fig.\ref{mvar}. 
Compared to our results,
(i) GWs in refs.\cite{Antusch:2016con, Antusch:2017vga, Amin:2018xfe} are based upon an inflation potential being different from our choice in eq.(\ref{V2}); 
(ii) GWs in ref.\cite{Lozanov:2019ylm} arise from a broader resonance with the mass scale in eq.(\ref{V2}) less than $M_P$ considered here;
and (iii) Ref.\cite{Garcia:2023eol} has only considered the GW for $n=4$, where the peak values of $k/m_{\rm{end}}\approx \{1.0, 1.5, 2.0,\cdots\}$ with $m_{\rm{end}}=\sqrt{3}\omega_{*}$ correspond to the peak values of $k/\omega_{*}\approx \{1.8, 2.7, 3.5,\cdots\}$ in the left plot of fig.\ref{mGW} respectively.

Finally, consider the individual GW spectrum in fig.\ref{mGWf} as radiation.
It contributes to the effective neutrino number $\Delta N_{\rm{eff}}\approx 1.16\times 10^{-5}~(7.78\times 10^{-7})$ for $n=4~(6)$ after substituting the results of fig.\ref{mGWf} into eq.(\ref{GWtot1}).

\section{Conclusion}
\label{con}
In this work we have forecasted GWs from preheating hosting GDM as the indirect probe of such GDM. 
Explicitly, we have considered two preheating models,
where in the Higgs (minimal) preheating the GWs are emitted by the Higgs (inflaton fluctuation) excitations due to the parameter (self-) resonance.
Our results show that the emitted GW spectra have distinct distributions in different high frequency ranges.
For the Higgs scalar excitations in the Higgs preheating, they give rise to the magnitudes of GW energy density spectra of order $10^{-10}$ at frequencies $10-10^{3}$ MHz depending on the GDM mass,
whereas for the inflaton fluctuation excitations in the minimal preheating they yield the magnitudes of GW energy density spectrum up to $10^{-9}~(10^{-11})$ at frequencies near $30~(2)$ MHz for the index $n=4~(6)$ with respect to the GDM mass of $1.04~(2.66)\times 10^{14}$ GeV.

While GWs at frequencies $10-10^{3}$ Hz have been detected by ground-based laser interferometers such as aLIGO \cite{aLIGO:2020wna} and GWs at frequencies $10^{-4}-1$ Hz are planned to be detected by space-based interferometers such as LISA \cite{LISACosmologyWorkingGroup:2022jok}, 
GW detectors at frequencies higher than $\sim$ MHz are lacking so far.
Over the years several concepts \cite{Aggarwal:2020olq} of high-frequency GW detectors have been proposed.
Despite being out of sensitivity of the next-generation CMB measurements on $\Delta N_{\rm{eff}}$ \cite{Abazajian:2019eic},
the derived GW spectra could be in the reaches of some of these proposals such as \cite{Herman:2022fau} in the near future.

There are several aspects left for future study. 
First, our idea of using GWs as the indirect probe of GDM from preheating can be also applied to GDM from reheating, 
although more diverse GW sources are expected in the later situation.
Second, the dependences of GW spectra on the lattice size $N$ and cut-off $k_{IR}$ can be further suppressed by more powerful compute sources.
Lastly, 
in order to eliminate the uncertainty in deriving the present-day GW spectra a more complete treatment on the equation of state between the end of preheating and of reheating than in the Appendix  is needed.

\appendix
\numberwithin{equation}{section}

\section{Evaluating gravitational wave spectrum}
\label{evalGW}
In this appendix we outline the evaluation of GW spectra from the end time of preheating $t_{*}$ to the present time $t_0$. 

First, the present-day frequency $f$ is related to the physical wavenumber $k_{*}(a_{*}/a_{\rm{end}})^{-1}$  at the end of preheating as 
\begin{eqnarray}\label{freq}
f=\frac{1}{2\pi}\left(\frac{a_{*}}{a_{0}}\right)\left(\frac{k_{*}}{a_{*}/a_{\rm{end}}}\right)=\frac{1}{2\pi}\left(\frac{a_{*}}{a_{\rm{rad}}}\frac{a_{\rm{rad}}}{a_{0}}\right)\left(\frac{a_{\rm{end}}}{a_{*}}\right)k_{*},
\end{eqnarray}
where the ratios of scale factors in the first bracket can be written as
\begin{eqnarray}\label{aratio}
\frac{a_{*}}{a_{\rm{rad}}}&=& \epsilon^{1/4}_{*}\left(\frac{\rho_{\rm{rad}}}{\rho_{*}}\right)^{1/4}\nonumber\\
\frac{a_{\rm{rad}}}{a_{0}}&=&G_{\rm{rad}}\left(\frac{\rho_{\rm{rad},0}}{\rho_{\rm{rad}}}\right)^{1/4},
\end{eqnarray}
with $\epsilon_{*}=(a_{*}/a_{\rm{rad}})^{1-3\bar{\omega}}$, $G_{\rm{rad}}=(g_{s,0}/g_{s,\rm{rad}})^{1/3}(g_{\rm{rad}}/g_{0})^{1/4}$ and $\rho_{*}\approx 3H^{2}_{*}M^{2}_{P}$.
Here, $\bar{\omega}$ is the equation of state between the end of preheating and the onset of radiation domination,
whereas $g_{s}$ and $g$ is the number of degrees of freedom associated with entropy and energy density respectively.
Plugging eq.(\ref{aratio}) into eq.(\ref{freq}) gives 
\begin{eqnarray}\label{freqf}
f&=&\frac{1}{2\pi}\left(\frac{1}{3}\right)^{1/4}\epsilon^{1/4}_{*}G_{\rm{rad}}\rho^{1/4}_{\rm{rad},0}\left(\frac{a_{\rm{end}}}{a_{*}}\right)\left(\frac{H_{*}}{M_{P}}\right)^{1/2}\left(\frac{k_{*}}{H_{*}}\right),\nonumber\\
~~&\approx& 5.5\times 10^{9}\left(\frac{a_{\rm{end}}}{a_{*}}\right)\left(\frac{H_{*}}{M_{P}}\right)^{1/2}\left(\frac{k_{*}}{H_{*}}\right)\rm{Hz},
\end{eqnarray}
where in the second line we have used $\epsilon_{*}\approx1$ as $\bar{\omega}$ has been close to $1/3$ at the end of preheating, 
$G_{\rm{rad}}\approx 0.788$, and $\rho_{\rm{rad},0}\approx 3.37\times 10^{-51}$ GeV$^{4}$. 

Second, the present-day GW spectrum is related to that at the end of preheating as
\begin{eqnarray}\label{GWs}
\Omega_{\rm{GW},0}(f)=\frac{1}{\rho_{c,0}}\left(\frac{d\rho_{\rm{GW}}}{dk}\right)_{0}=\left(\frac{a_{*}}{a_{0}}\right)^{4}\frac{\rho_{*}}{\rho_{c,0}}\Omega_{\rm{GW},*}(k)
\end{eqnarray}
where $\rho_{c,0}$ is the present critical energy density and $\rho_{\rm{GW}}$ scales as $\sim a^{-4}$.
Substituting the value of $a_{*}/a_{0}$ in terms of eq.(\ref{aratio}) into eq.(\ref{GWs})  one obtains
\begin{eqnarray}\label{GWsf}
\Omega_{\rm{GW},0}(f)&=&\epsilon_{*}G^{4}_{\rm{rad}}\Omega_{\rm{rad},0}\Omega_{\rm{GW},*}(k), \nonumber\\
~~&\approx& 2.0\times 10^{-5}\Omega_{\rm{GW},*}(k),
\end{eqnarray}
where  $k_{*}$ is replaced by $f$ via eq.(\ref{freqf}) and  in the second line $\Omega_{\rm{rad},0}\approx 5\times 10^{-5}$ has been used.
The main uncertainty in eq.(\ref{GWsf}) arises from $\epsilon_{*}$ which is sensitive to the value of $\bar{\omega}$.
Take $a_{*}/a_{\rm{rad}}\approx 10^{-5}$ for example, $\bar{\omega}=0.3~(0.32)$ gives $\epsilon_{*}=0.32~(0.63)$.

\end{document}